\documentclass[twocolumn]{fairmeta}
\usepackage{amsmath}
\usepackage{newtxtext,newtxmath}
\usepackage{cuted}
\usepackage{tabularx}
\usepackage{url}
\usepackage{listings}
\usepackage{tikz}
\usetikzlibrary{arrows.meta, calc, positioning, shapes.geometric, decorations.pathreplacing, backgrounds}
\usepackage{algorithm}
\usepackage{algpseudocode}
\usepackage{pgfplots}
\pgfplotsset{compat=1.18}
\usepackage{chemformula}
\usepackage[version=4]{mhchem}
\usepackage{float}
\usepackage{textcomp}

\usepackage[most]{tcolorbox}
\newtcolorbox{agentthought}{
  colback=activepurple!8, colframe=activepurple!60,
  fonttitle=\small\sffamily\bfseries, title={Agent thought ($t_i$)},
  boxrule=0.4pt, arc=2pt, left=4pt, right=4pt, top=2pt, bottom=2pt,
  before skip=4pt, after skip=4pt
}
\newtcolorbox{toolcall}[1][]{
  colback=ragblue!8, colframe=ragblue!60,
  fonttitle=\small\sffamily\bfseries, title={#1},
  boxrule=0.4pt, arc=2pt, left=4pt, right=4pt, top=2pt, bottom=2pt,
  before skip=4pt, after skip=4pt
}
\newtcolorbox{observation}{
  colback=psmilesgreen!8, colframe=psmilesgreen!60,
  fonttitle=\small\sffamily\bfseries, title={Observation ($o_i$)},
  boxrule=0.4pt, arc=2pt, left=4pt, right=4pt, top=2pt, bottom=2pt,
  before skip=4pt, after skip=4pt
}

\definecolor{deepblue}{RGB}{0,0,120}
\definecolor{midgreen}{RGB}{0,120,0}
\definecolor{darkred}{RGB}{120,0,0}
\definecolor{codecolor}{RGB}{240,240,240}
\definecolor{ragblue}{RGB}{51,122,183}
\definecolor{psmilesgreen}{RGB}{92,184,92}
\definecolor{mdorange}{RGB}{240,173,78}
\definecolor{activepurple}{RGB}{150,100,200}
\definecolor{expbteal}{RGB}{0,128,128}
\definecolor{expccopper}{RGB}{180,90,40}

\titleformat{\section}{\large\sffamily\bfseries}{\thesection}{0.5em}{}
\titleformat{\subsection}{\normalsize\sffamily\bfseries}{\thesubsection}{0.5em}{}
\titleformat{\subsubsection}{\normalsize\sffamily\bfseries}{\thesubsubsection}{0.5em}{}
\titlespacing*{\section}{0pt}{0.85ex plus 0.2ex minus 0.1ex}{0.35ex plus 0.1ex}
\titlespacing*{\subsection}{0pt}{0.7ex plus 0.15ex}{0.3ex plus 0.1ex}
\titlespacing*{\subsubsection}{0pt}{0.6ex plus 0.1ex}{0.28ex plus 0.05ex}

\title{Towards Discovery of Polymers for Insulin Delivery via Physics-Grounded Agentic Workflows}

\author[1]{Martins Otun}
\affiliation[1]{Algonix AI Ltd., Scotland, United Kingdom}
\metadata[Code]{\url{https://github.com/otunmartins/FRIDGEFREENET}}
\correspondence{O.M. (\email{martins@algonix.co.uk})}

\abstract{%
Cold-chain storage limits access to insulin for hundreds of millions of people; a thermally protective patch polymer could help, but the design space is too large for exhaustive experiment.
Starting from that problem, we narrow to an agentic workflow: a large language model (LLM) calls physics-based tools through the Model Context Protocol (MCP), searching the discrete PSMILES space under a budget of OpenMM Packmol-matrix evaluations.
The LLM acts as an implicit acquisition function conditioned on a persistent ``discovery world'': hypotheses, literature claims, and simulation outcomes updated each iteration.
Under matched oracle budgets, the best autonomous campaign reaches an insulin--polymer interaction energy of $-2263$~kJ/mol, outperforming reinforcement-learning baselines by 68\% and Bayesian optimization by 19\%.
Three independent campaigns converge on one structural motif (dense hydrogen-bond donors and acceptors per repeat unit) while physics checks reject infeasible packings and name--structure mismatches before they steer the next step.
The science stage is CPU-bound and runs on commodity hardware.
More broadly, the same architecture and workflow designed here applies to other protein-stabilization tasks whenever a tractable screening oracle is available.}

\begin{document}

\maketitle

\section{Introduction}

Diabetes affects over 500 million people worldwide, and its management depends critically on insulin that must be stored between 2 and 8$^{\circ}$C~\cite{hsu2022glycosylated}.
In resource-limited settings, where refrigeration is unreliable or absent, this cold-chain requirement leads to degraded medication and preventable complications.
A transdermal patch whose polymer matrix stabilizes insulin at ambient temperature would fundamentally change access to treatment, but the combinatorial space of candidate polymers is vast, and no single experimental campaign can screen more than a tiny fraction of it.

Polymer informatics has begun to narrow this space computationally.
Machine-readable repeat-unit notations such as PSMILES~\cite{lin2019bigsmiles,kuenneth2023polybert} and graph-neural-network property predictors trained on curated databases~\cite{huan2020polymer,kern2022solvent,gurnani2024ai} enable rapid surrogate evaluation.
However, surrogate models approximate physical reality rather than simulate it; they cannot catch sterically infeasible packings or reject chemically implausible structures that happen to score well on a learned surface.
What is missing is a closed loop in which candidate generation, explicit molecular simulation, and hypothesis revision operate together so that each iteration is grounded in physics rather than interpolation.

Meanwhile, large language model (LLM) agents have demonstrated the ability to orchestrate multi-step scientific workflows.
Boiko et al.\ showed autonomous chemical synthesis planning~\cite{boiko2023autonomous}, Zheng et al.\ applied ChatGPT to metal--organic framework prediction~\cite{zheng2023chatgpt}, and the Kosmos system combined parallel agents through a shared world model for multi-domain discovery~\cite{kosmos2025}.
Stephens and Salawu formalized such agentic processes as probability chains over tool calls, providing a mathematical vocabulary for analyzing how state updates, inference functionals, and action-space partitioning affect the likelihood of reaching a goal~\cite{stephens2025mathframing}.
Those examples emphasize synthesis planning, porous-crystal prediction, or cross-domain literature agents rather than budgeted search over polymer repeat units with explicit energy evaluation against a specified protein structure between hypothesis updates.
Complementary work increasingly connects LLMs to molecular dynamics and simulation stacks, including simulation-informed training of MD assistants~\cite{mdagent2_2026}, natural-language control of MD engines~\cite{simulon2026}, and autonomous simulation agents for polymer benchmark workflows~\cite{liu2024asa}.

In this work, we focus on a different closed loop that combines literature mining, cheminformatics, and molecular simulations to screen a range of polymeric materials. We instantiate this design as an agentic discovery platform built on the Model Context Protocol (MCP)~\cite{mcp2024}.
The platform exposes literature mining, PSMILES validation, cheminformatics mutation, and OpenMM Packmol-matrix screening as callable tools; a persistent ``discovery world'' accumulates hypotheses, simulation outcomes, and literature claims across iterations.
Formally, we cast the search as budget-constrained optimization over the discrete PSMILES space, where the LLM serves as an implicit acquisition function~\cite{yang2024opro} conditioned on a structured state that classical optimizers lack~\cite{shahriari2016bayesopt}.
The simulation stage is CPU-bound, so the entire workflow runs on commodity hardware, compatible with the local and edge LLM deployments gaining traction in resource-constrained laboratories~\cite{xu2024ondevice}.

The paper is organized as follows.
First, we formalize the agentic discovery loop using the probabilistic-chain framework of Stephens and Salawu and budget-constrained regret theory, and we compare the degrees of freedom available to RL, Bayesian optimization, and LLM-agent strategies (\S\ref{sec:methods}).
Second, we benchmark DQN, PPO, and Optuna TPE against three independent 25-iteration autonomous LLM campaigns on the same physics objective, showing that the agentic workflow reduces simple regret by up to 68\% over RL (\S\ref{sec:results}).
Third, we analyze the chemical reasoning recorded in the discovery world, demonstrating cross-campaign convergence on hydrogen-bond-dense motifs and quantifying how physics-based falsification prunes hallucinated or sterically untenable proposals (\S\ref{sec:analysis}).

\section{Methods}
\label{sec:methods}

\subsection{Platform architecture}

We implement the platform as a single Python MCP server~\cite{mcp2024} (FastMCP) that registers tools for literature mining, polymer validation, molecular screening, cheminformatics mutation, and session management.
It runs inside the \texttt{insulin-ai-sim} conda environment, which bundles OpenMM~\cite{eastman2017openmm}, RDKit~\cite{landrum2013rdkit}, Packmol~\cite{packmol2009}, the Open Force Field toolkit~\cite{qiu2021development}, and Stable-Baselines3~\cite{sb3_2021}.
An LLM hosted by OpenCode (or Cursor) connects to this server and calls tools in response to user prompts; model weights, tool schema, and session artifacts are version-controlled alongside the scientific code.

\begin{figure*}[t]
\centering
\begin{tikzpicture}[
    font=\small,
    box/.style={rectangle, rounded corners=3pt, draw, thick,
                minimum width=2.35cm, minimum height=0.92cm, align=center},
    litbox/.style={box, fill=ragblue!15, draw=ragblue},
    genbox/.style={box, fill=psmilesgreen!15, draw=psmilesgreen},
    simbox/.style={box, fill=mdorange!15, draw=mdorange},
    statebox/.style={box, fill=activepurple!15, draw=activepurple,
                     minimum width=8.6cm, minimum height=0.88cm},
    topbox/.style={box, fill=white, minimum width=2.9cm, minimum height=0.88cm},
    arr/.style={-{Stealth[length=2.2mm]}, thick},
    fb/.style={arr, dashed, color=red!70},
    lbl/.style={font=\scriptsize, inner sep=1pt, fill=white, fill opacity=0.92, text opacity=1}
]
\node[genbox] (gen) {PSMILES\\generate + validate};
\node[litbox, left=2.35cm of gen] (lit) {Literature\\mining};
\node[simbox, right=2.35cm of gen] (sim) {OpenMM\\Packmol screen};
\node[topbox, above=0.58cm of gen] (llm) {LLM agent\\{\scriptsize OpenCode\,/\,Cursor}};
\node[statebox, below=0.58cm of gen] (world) {Discovery world + iteration state};

\coordinate (genTopIn) at ($(gen.north west)!0.22!(gen.north east)$);
\coordinate (genBotOut) at ($(gen.south west)!0.78!(gen.south east)$);
\coordinate (worldTopIn) at (genBotOut |- world.north);

\draw[arr] (llm.south west) -- (lit.north);
\draw[arr] ($(llm.south west)!0.42!(llm.south east)$) -- (genTopIn);
\draw[arr] (llm.south east) -- (sim.north);

\draw[arr] (lit.east) -- (gen.west) node[lbl, midway, above=2pt] {candidates};
\draw[arr] (gen.east) -- (sim.west) node[lbl, midway, above=2pt] {PSMILES};

\draw[arr] (genBotOut) -- (worldTopIn)
  node[lbl, pos=0.52, anchor=east, xshift=-5pt, inner sep=2pt] {patch / query};

\def\fbdrop{0.16}
\def\fbin{0.15}
\draw[fb] (sim.south) -- ++(0,-\fbdrop) -| ($(world.north east)+(-\fbin,0)$);
\draw[fb] ($(world.north west)+(\fbin,0)$) -| (lit.south);
\end{tikzpicture}
\caption{Platform architecture. Solid arrows denote tool calls from the LLM agent; dashed arrows show feedback paths through the session state. The discovery world file persists hypotheses, literature claims, and simulation outcomes for cross-iteration planning.}
\label{fig:architecture}
\end{figure*}
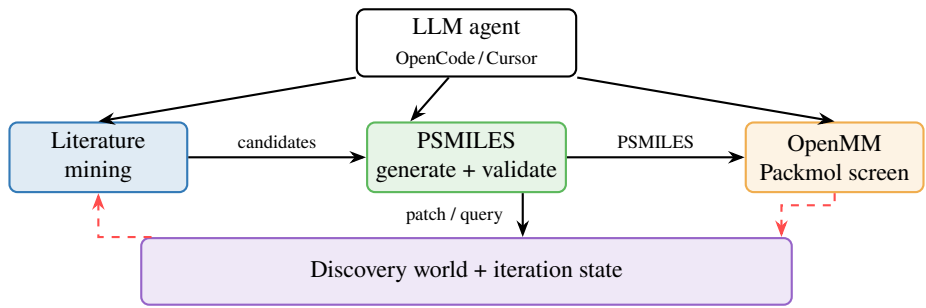

Figure~\ref{fig:architecture} summarizes the data flow.
Each discovery session creates a folder under \texttt{runs/<session\_id>/} containing per-iteration state files (\texttt{agent\_iteration\_N.json}), a structured world-model rollup (\texttt{discovery\_world.json}), cumulative best-candidate records (\texttt{ALL\_ITERATIONS\_BEST\_CANDIDATES.tsv}), structure artifacts (PDB, PNG), and a compiled summary report (Markdown and PDF).

\subsection{Problem formulation}
\label{sec:formulation}

We formalize the agentic loop at two complementary levels: a \emph{probabilistic chain} that describes the agent's action sequence~\cite{stephens2025mathframing}, and a \emph{budget-constrained optimization} that defines the search objective~\cite{shahriari2016bayesopt}.

\paragraph{Probabilistic chain (agent level).}
Following Stephens and Salawu~\cite{stephens2025mathframing}, we model the discovery loop as a chain of conditional probabilities over tool calls.
Given an initial context $c$ (user objective, tool schema, and seed state), the probability of executing an action sequence $\mathbf{a} = (a_1, \ldots, a_n)$ is
\begin{equation}
P(\mathbf{a} \mid c) = \prod_{i=1}^{n} P(a_i \mid s_{i-1},\, c),
\label{eq:chain}
\end{equation}
where $s_0 = s_0(c)$ is the initial prompt state and $s_i = u(a_i, s_{i-1})$ is the state update~\cite{puterman1994mdp}.
Under the ReAct formalism the LLM generates a reasoning trace (``thought'') $t_i$ before each tool call, so
\begin{equation}
P^t(a_i \mid s_{i-1}) = P(a_i \mid t_i, s_{i-1})\; P(t_i \mid s_{i-1}).
\label{eq:react}
\end{equation}
Each action decomposes as $a_i \to (\alpha_i, x_i, o_i)$, where $\alpha_i \in \{\texttt{mine},\, \texttt{generate},\, \texttt{mutate},\, \texttt{screen},\, \texttt{patch\_world}\}$ is the MCP tool class, $x_i$ is its argument, and $o_i$ is the returned observation.

Different search strategies occupy different positions in this framework depending on the optimization levers (the \emph{degrees of freedom}) available to each (Table~\ref{tab:dof}).
RL agents use a fixed policy network $\mathcal{F}$, no persistent state update ($\mathcal{W}_t = \emptyset$), and a small mutation library $\{\alpha\}$.
Optuna adds an explicit acquisition function (tree-structured Parzen estimator) but still lacks hypothesis-level reasoning.
The LLM agent has access to all levers: an inference functional $\mathcal{F}$ that improves through in-context learning, a structured selective state update $u_\text{selective}$ (the discovery world), and the full MCP tool set.
Per Stephens and Salawu~\cite{stephens2025mathframing}, MCP provides standardized tool schemas and argument syntax that constrain the action space, improving $P^{\mathcal{F},u}(a_i \mid s_{i-1})$ relative to unconstrained generation.

\begin{table}[t]
\centering
\caption{Degrees of freedom available to each search method, following the probabilistic-chain framework of Stephens and Salawu~\cite{stephens2025mathframing}.}
\label{tab:dof}
\small
\begin{tabular}{lcccc}
\toprule
 & $\mathcal{F}$ & $u$ & $s_0(c)$ & $\{\alpha\}$ \\
\midrule
RL    & policy net & $\emptyset$ & PEG seed  & mutations \\
TPE   & sampler   & $\emptyset$ & seed      & mutations \\
LLM   & $\pi_\text{LLM}$ & $u_\text{sel}$ & prompt+world & MCP tools \\
\bottomrule
\end{tabular}
\end{table}

\paragraph{Budget-constrained optimization (search level).}
Let $\mathcal{X}$ denote the discrete space of valid PSMILES and let $f\!:\!\mathcal{X}\!\to\!\mathbb{R}$ be the noisy screening oracle that returns the interaction energy,
\begin{equation}
f(x) = E_\text{int}(x) + \epsilon, \quad \epsilon \sim \mathcal{N}(0,\,\sigma^2),
\label{eq:oracle}
\end{equation}
where $\sigma$ captures stochastic packing variability (${\sim}$15\% in our protocol).
The agent selects candidates $x_1,\ldots,x_B$ under oracle budget $B$; the goal is to minimize the \emph{simple regret}~\cite{shahriari2016bayesopt,srinivas2010gpucb}
\begin{equation}
r_B = f(x^\star) - \min_{t \le B} f(x_t).
\label{eq:regret}
\end{equation}
Classical Bayesian optimization chooses the next query by maximizing an explicit acquisition function $\alpha(x \mid \mathcal{D}_t)$ such as expected improvement or GP-UCB~\cite{srinivas2010gpucb}.
The LLM agent instead implements an \emph{implicit acquisition}~\cite{yang2024opro,liu2025igpo},
\begin{equation}
x_{t+1} = \pi_\text{LLM}(s_t) \;\approx\; \arg\max_{x\in\mathcal{X}}\; \alpha_\text{implicit}(x \mid \mathcal{D}_t, \mathcal{W}_t),
\label{eq:implicit_acq}
\end{equation}
conditioned on the full state $s_t = (\mathcal{D}_t, \mathcal{W}_t, \mathcal{H}_t)$, which includes the evaluation history $\mathcal{D}_t$, the discovery-world hypotheses $\mathcal{W}_t$, and the LLM's implicit belief $\mathcal{H}_t$.
Iterations that \emph{falsify} a hypothesis (packing failures, name--structure mismatches) contribute high information gain
\begin{equation}
I_t = H(\mathcal{H}_t) - H(\mathcal{H}_t \mid o_t,\, \mathcal{W}_{t+1})
\label{eq:infogain}
\end{equation}
even when $f(x_t)$ is poor.
Such observations are wasted budget under RL or Optuna (which have no hypothesis state $\mathcal{W}_t$) but valuable in the agentic loop because they prune unpromising chemical families from future exploration.

\subsection{Discovery world model}

Inspired by the structured world model in Kosmos~\cite{kosmos2025}, we maintain a session-scoped JSON document (\texttt{discovery\_world.json}) that accumulates:
\begin{itemize}
\item \textbf{Objective and history:} the user's research goal and any mid-campaign steering notes.
\item \textbf{Literature entries:} stable IDs, titles, one-sentence claims, source iteration, and optional DOI or Semantic Scholar identifiers.
\item \textbf{Simulation entries:} PSMILES, interaction energy, screening status, and pointers to structure artifacts.
\item \textbf{Hypotheses:} free-text statements with supporting evidence IDs and a status tag (open, supported, weak, dropped).
\item \textbf{Open questions and human directives:} recorded per iteration for traceability.
\end{itemize}
Three MCP tools operate on this file: \texttt{patch\_discovery\_world} merges id-keyed lists (update existing or append new entries); \texttt{discovery\_world\_planning\_context} returns a bounded text digest for prompt construction; and \texttt{get\_discovery\_world\_state} provides full or summarized JSON for inspection.
When the agent calls \texttt{save\_discovery\_state} after each iteration, the server automatically updates \texttt{meta.last\_iteration} and links the latest \texttt{agent\_iteration\_N.json} into the world file, ensuring the rollup stays synchronized without requiring the model to issue a second patch.

\subsection{Agentic discovery loop}

A single iteration proceeds as follows:
\begin{enumerate}
\item \textbf{Literature mining.} The agent queries Semantic Scholar~\cite{lo2020s2orc} (or an Asta-backed corpus when an API key is available) with an adaptive query derived from the world model's hypotheses, open questions, and prior top performers.
\item \textbf{Candidate generation.} Material names from the literature are converted to PSMILES via a curated 60-polymer lookup table with PubChem fallback (automated polymerization-site detection for vinyl, ester, and amide monomers). Each candidate is validated with RDKit (connectivity, valence, exactly two \texttt{[*]} connection points) and cross-checked against PubChem monomer Tanimoto similarity.
\item \textbf{OpenMM screening.} Validated PSMILES are evaluated by the Packmol-matrix protocol: short polymer oligomers are packed around insulin (PDB 4F1C), the composite is energy-minimized with the GAFF/OpenFF force field for polymers and AMBER ff14SB for the protein~\cite{he2020amber,maier2015ff14sb,qiu2021development}, and the insulin--polymer interaction energy is computed.
\item \textbf{Mutation.} A cheminformatics mutator generates derivative PSMILES from top performers via functional-group substitution, backbone extension, and random fragment recombination, optionally guided by feedback on problematic substructures.
\item \textbf{State persistence.} The agent saves iteration state, patches the world model with new literature and simulation entries, updates hypotheses, and appends the best candidate to a cumulative TSV.
\end{enumerate}
In \emph{human-in-the-loop} mode the agent pauses after step~5 for user feedback; in \emph{autonomous} mode it immediately proceeds to the next iteration, stopping after $N$ cycles or upon energy-threshold or saturation criteria.

Algorithm~\ref{alg:agentic} formalizes this loop using the notation from \S\ref{sec:formulation}.

\begin{algorithm}[t]
\caption{Agentic Discovery Loop}\label{alg:agentic}
\small
\begin{algorithmic}[1]
\Require context $c$, oracle $f$, budget $B$
\State $s_0 \gets s_0(c)$ \Comment{prompt + tool schema}
\State $\mathcal{W}_0 \gets \emptyset$; $\mathcal{D}_0 \gets \emptyset$
\For{$t = 1, \ldots, B$}
  \State $t_i \gets \text{LLM reason over } s_{t-1}$ \Comment{Eq.~\ref{eq:react}: thought}
  \State $\alpha_i \gets$ select MCP tool class from $\{\texttt{mine}, \texttt{gen}, \texttt{mut}, \texttt{scr}\}$
  \State $x_i \gets$ LLM generate tool argument
  \State $o_i \gets \alpha_i(x_i)$ \Comment{execute tool call}
  \If{$\alpha_i = \texttt{screen}$}
    \State $y_t \gets f(x_i)$ \Comment{Eq.~\ref{eq:oracle}: noisy oracle}
    \State $\mathcal{D}_t \gets \mathcal{D}_{t-1} \cup \{(x_i, y_t)\}$
  \EndIf
  \State $s_t \gets u_\text{sel}(a_i,\, s_{t-1})$ \Comment{Eq.~\ref{eq:chain}: state update}
  \State $\mathcal{W}_t \gets \texttt{patch\_world}(\mathcal{W}_{t-1},\, o_i,\, t_i)$
\EndFor
\State \Return $\arg\min_{x \in \mathcal{D}_B} f(x)$
\end{algorithmic}
\end{algorithm}

\subsection{Molecular screening protocol}
\label{sec:screening}

The screening objective is the non-bonded interaction energy between insulin and a polymer shell.
For each candidate PSMILES, the server:
(a) generates a 3D oligomer conformer with RDKit;
(b) packs $n$ polymer chains around insulin using Packmol~\cite{packmol2009} at a target density;
(c) assigns GAFF2/OpenFF parameters to the polymer and AMBER ff14SB to insulin;
(d) runs a two-stage energy minimization (steepest descent followed by L-BFGS);
(e) optionally performs a short NPT equilibration at 310~K and 1~bar using the LangevinMiddleIntegrator~\cite{eastman2017openmm} with a 2~fs timestep.
The interaction energy $E_\text{int}$ is defined as
\begin{equation}
E_\text{int} = E_\text{complex} - E_\text{insulin} - E_\text{polymer}
\label{eq:eint}
\end{equation}
where each term is the potential energy of the specified subsystem after minimization.
More negative values indicate stronger insulin--polymer attraction.
Each call to \texttt{evaluate\_candidates} corresponds to one oracle query $f(x_t)$ (Eq.~\ref{eq:oracle}), returning observation $o_t = (x_t, y_t, \text{status}_t)$.
This single-point heuristic does not replace free-energy calculations but provides a computationally tractable ranking across hundreds of candidates.

Candidates that fail RDKit parsing, violate prescreen chemistry rules (e.g.\ forbidden elements, excessive molecular weight), or exceed Packmol packing timeouts are logged with verbatim failure reasons and excluded from energy rankings.

\subsection{Benchmark methods}
\label{sec:benchmark_methods}

To contextualize the agentic workflow, we benchmark three non-agentic search strategies under a matched evaluation budget of $B{=}160$ successful OpenMM screenings.

\paragraph{DQN and PPO reinforcement learning.}
We adapt an RL formulation~\cite{boiko2023autonomous} to polymer discovery.
In the notation of \S\ref{sec:formulation}, both agents operate on a degenerate state $s_t = (x_t^\text{best}, y_t^\text{best})$ with no world model ($\mathcal{W}_t = \emptyset$), a fixed policy network as inference functional $\mathcal{F}$, and the mutation library as their only action set $\{\alpha\}$ (Table~\ref{tab:dof}).
Actions select from functional-group swap, backbone extension, and fragment insertion; the reward is $-f(x_t)$.
We train DQN and PPO agents using Stable-Baselines3~\cite{sb3_2021} for 160 timesteps (20 episodes $\times$ 8 steps), seeding from \texttt{[*]OCC[*]} (polyethylene glycol repeat unit).
Each agent is run with three random seeds (42, 123, 456) to estimate variance.

\paragraph{Optuna TPE.}
Optuna's tree-structured Parzen estimator (TPE)~\cite{optuna2019} implements an explicit acquisition function (the density ratio $p(x \mid y < y^\star) / p(x)$) but without the hypothesis-level reasoning $\mathcal{W}_t$ or structured state update that the LLM agent employs.
It searches over mutator seed and feedback fraction, generating 8 PSMILES candidates per trial across 20 trials.
The objective maximizes a composite discovery score that rewards low interaction energy and chemical diversity.

\paragraph{Evaluation protocol.}
All methods use the same \texttt{MDSimulator.evaluate\_candidates} entry point, the same insulin PDB (4F1C), the same force field assignments, and \texttt{INSULIN\_AI\_EVAL\_MAX\_WORKERS=1} for deterministic execution order.
Results are recorded in a shared TSV schema with columns for method, number of evaluations, best interaction energy, unique PSMILES evaluated, and wall-clock time.

\section{Benchmark Results}
\label{sec:results}

We executed all OpenMM screenings on CPU without GPU acceleration.
Table~\ref{tab:benchmarks} summarizes performance across all methods.

\begin{table*}[t]
\centering
\caption{Benchmark and agentic results. RL rows: mean $\pm$ std over seeds 42, 123, 456. Optuna: seed~42. Campaign session identifiers appear in Table footnotes; $N_\text{eval}$ for Campaign~C estimated from session transcript (${\sim}$50+ screened, ${\sim}$10 Packmol failures).}
\label{tab:benchmarks}
\small
\begin{tabular}{lccccc}
\toprule
\textbf{Method} & \textbf{$N_\text{eval}$} & \textbf{$E_\text{int}^\text{best}$ (kJ/mol)} & \textbf{Unique PSMILES} & \textbf{Wall time} & \textbf{Best candidate} \\
\midrule
DQN (3 seeds)   & $99 \pm 1$  & $-1345 \pm 67$   & 14 & ${\sim}$8 min  & \texttt{[*]CCOC([*])=O} \\
PPO (3 seeds)   & $100$       & $-1314 \pm 185$  & 14 & ${\sim}$8 min  & \texttt{[*]CCOC([*])=O} \\
Optuna TPE      & 143         & $-1902$          & 26 & ${\sim}$1.9 h  & \texttt{[*]NC(=O)NC([*])=O} \\
Campaign~A (LLM)   & 115+        & $\mathbf{-2263}$ & 25+ & ${\sim}$6 h  & \texttt{[*]CNC(=O)NC([*])=O} \\
Campaign~B (LLM)   & 30          & $-1545$          & 30 & ${\sim}$4 h  & polyhistidine \\
Campaign~C (LLM)   & 50+         & $-1765$          & 50+ & ${\sim}$5 h  & polygalacturonic acid \\
\bottomrule
\end{tabular}
\end{table*}

\subsection{Baseline performance (RL and Optuna)}

Both RL algorithms converge rapidly (the first mutation from PEG already produces a favorable candidate) but plateau early.
DQN achieves a cross-seed mean of $-1345 \pm 67$~kJ/mol; PPO reaches $-1314 \pm 185$~kJ/mol with wider variance, reflecting greater sensitivity to the random seed under on-policy updates.
Each method evaluates ${\sim}$100 candidates in ${\sim}$450~s but discovers only 14 unique PSMILES, indicating substantial re-evaluation.
The fixed mutation library confines both agents to ester and ether backbones; neither discovers amide-rich or nitrogen-heterocyclic motifs.
In the language of \S\ref{sec:formulation}, RL operates with $\mathcal{W}_t = \emptyset$ and a fixed $\{\alpha\}$, so it cannot escape the chemical neighborhood of its PEG seed.

Optuna evaluates 143 candidates across 20 trials, discovering 26 unique PSMILES and reaching $-1902$~kJ/mol, a substantial improvement over RL at 15$\times$ the wall time (6807~s).
TPE's explicit acquisition function samples more broadly, finding the urea-bridged structure \texttt{[*]NC(=O)NC([*])=O} ($-1683$~kJ/mol) that RL never encounters.
However, it also proposes a hydrolytically unstable peroxide-ester (\texttt{[*]OOC(=O)C([*])=O}, $-1171$~kJ/mol), illustrating energy-only optimization without chemical judgment.

\subsection{Agentic campaigns}

We ran three independent 25-iteration LLM campaigns, each with a different exploration strategy.

\textbf{Campaign~A} emphasizes synthetic repeat units, heterocyclic urethanes, and mutation-guided refinement.
It evaluates over 115 candidates and yields three scaffolds below $-2100$~kJ/mol, with best energy $-2263$~kJ/mol for poly(N-acetyl-1,2-diaminoethane), a 19\% gain over Optuna and 68\% over DQN (Figure~\ref{fig:running_best}).
The running-best curve rises through oxazolidinone and lactamide intermediates to the C2 amide breakthrough at iteration~15, with hydroxamic and hydrazide analogs extending the frontier through iteration~25.

\textbf{Campaign~B} starts from a curated polymer table (PEG, PLGA, chitosan), then pivots after mid-campaign stagnation toward amino acid polymers under a biomimetic hypothesis.
Only 30 distinct PSMILES are evaluated, yet the running best improves from chitosan ($-1460$~kJ/mol) to polyhistidine ($-1545$~kJ/mol) at iteration~23.
Campaign~B explores a different subspace ($\alpha$-amino acid backbones) and independently reinforces the importance of dense hydrogen-bonding groups.

\textbf{Campaign~C} emphasizes acidic polysaccharides and uronic acids.
Polygalacturonic acid reaches $-1765$~kJ/mol at iteration~7 (Table~\ref{tab:agentic_top_c}), with polyglucuronic acid ($-1574$~kJ/mol) and polyhydrazide ($-1496$~kJ/mol) as runners-up.
The campaign screened approximately 50+ candidates, with ${\sim}$10 Packmol failures (high-charge polylysine, aromatic polytyrosine) paralleling the physics-driven rejection observed in Campaigns~A and B.

\begin{figure*}[t]
\centering
\begingroup
\fontfamily{cmr}\selectfont
\pgfplotsset{
  cmstyle/.style={
    tick label style={font=\fontsize{7}{8}\fontfamily{cmr}\selectfont,
                      /pgf/number format/fixed},
    label style={font=\fontsize{7.5}{9}\fontfamily{cmr}\selectfont},
    title style={font=\fontsize{7.5}{9}\fontfamily{cmr}\selectfont},
    grid=major,
    grid style={gray!25},
    mark options={solid},
    every axis/.append style={thick}
  }
}
\subcaptionbox{Running-best $E_\text{int}$ vs.\ iteration.\label{fig:running_best}}{%
\begin{tikzpicture}
\begin{axis}[
    cmstyle,
    width=0.48\textwidth,
    height=5.4cm,
    xlabel={Iteration},
    ylabel={Running-best $E_{\mathrm{int}}$ (kJ/mol)},
    xmin=0, xmax=26,
    ymin=-2400, ymax=-200,
    legend style={at={(0.5,1.03)}, anchor=south,
                  legend columns=3,
                  font=\fontsize{6}{7}\fontfamily{cmr}\selectfont,
                  draw=gray!40, fill=white, fill opacity=0.92,
                  /tikz/every even column/.append style={column sep=3pt}},
]
\addplot[activepurple, thick, mark=*, mark size=1pt] coordinates {
    (1,-1450) (2,-1473) (3,-1549) (4,-1549) (5,-1549)
    (6,-1549) (7,-1549) (8,-1665) (9,-1665) (10,-1665)
    (11,-1665) (12,-1807) (13,-1807) (14,-1807) (15,-2263)
    (16,-2263) (17,-2263) (18,-2263) (19,-2263) (20,-2263)
    (21,-2263) (22,-2263) (23,-2263) (24,-2263) (25,-2263)
};
\addplot[expbteal, thick, mark=square*, mark size=1.2pt] coordinates {
    (1,-1460) (2,-1460) (3,-1460) (4,-1460) (5,-1460)
    (6,-1460) (7,-1460) (8,-1460) (9,-1460) (10,-1460)
    (11,-1460) (12,-1460) (13,-1460) (14,-1460) (15,-1460)
    (16,-1460) (17,-1460) (18,-1460) (19,-1460) (20,-1460)
    (21,-1460) (22,-1460) (23,-1545) (24,-1545) (25,-1545)
};
\addplot[expccopper, thick, mark=triangle*, mark size=1.3pt] coordinates {
    (1,-1370) (2,-1370) (3,-1385) (4,-1385) (5,-1385)
    (6,-1385) (7,-1765) (8,-1765) (9,-1765) (10,-1765)
    (11,-1765) (12,-1765) (13,-1765) (14,-1765) (15,-1765)
    (16,-1765) (17,-1765) (18,-1765) (19,-1765) (20,-1765)
    (21,-1765) (22,-1765) (23,-1765) (24,-1765) (25,-1765)
};
\addplot[ragblue, dashed, thick, domain=0:26] {-1345};
\addplot[psmilesgreen, dotted, thick, domain=0:26] {-1314};
\addplot[mdorange, dashdotted, thick, domain=0:26] {-1902};
\legend{Camp.\ A, Camp.\ B, Camp.\ C, DQN mean, PPO mean, Optuna best}
\end{axis}
\end{tikzpicture}%
}\hfill
\subcaptionbox{Best $E_\text{int}$ vs.\ unique PSMILES evaluated.\label{fig:comparison}}{%
\begin{tikzpicture}
\begin{axis}[
    cmstyle,
    width=0.48\textwidth,
    height=5.4cm,
    xlabel={Unique PSMILES evaluated},
    ylabel={Best $E_{\mathrm{int}}$ (kJ/mol)},
    xmin=8, xmax=58,
    ymin=-2400, ymax=-900,
]
\addplot[only marks, mark=square*, mark size=2.5pt, ragblue,
         error bars/.cd, y dir=both, y explicit]
    coordinates {(14, -1345) +- (0, 67)};
\addplot[only marks, mark=triangle*, mark size=3pt, psmilesgreen,
         error bars/.cd, y dir=both, y explicit]
    coordinates {(14, -1314) +- (0, 185)};
\addplot[only marks, mark=*, mark size=2.5pt, mdorange]
    coordinates {(26, -1902)};
\addplot[only marks, mark=diamond*, mark size=3.5pt, activepurple]
    coordinates {(25, -2263)};
\addplot[only marks, mark=pentagon*, mark size=3.5pt, expbteal]
    coordinates {(30, -1545)};
\addplot[only marks, mark=+, mark size=3.5pt, expccopper,
         mark options={thick,scale=1.6}]
    coordinates {(50, -1765)};
\node[font=\fontsize{6}{7}\fontfamily{cmr}\selectfont, anchor=south west]
    at (axis cs:15.5,-1278) {DQN};
\node[font=\fontsize{6}{7}\fontfamily{cmr}\selectfont, anchor=south west]
    at (axis cs:15.5,-1100) {PPO};
\node[font=\fontsize{6}{7}\fontfamily{cmr}\selectfont, anchor=east]
    at (axis cs:24,-1902) {Optuna};
\node[font=\fontsize{6}{7}\fontfamily{cmr}\selectfont, anchor=west]
    at (axis cs:26.5,-2263) {A};
\node[font=\fontsize{6}{7}\fontfamily{cmr}\selectfont, anchor=west]
    at (axis cs:31.5,-1545) {B};
\node[font=\fontsize{6}{7}\fontfamily{cmr}\selectfont, anchor=west]
    at (axis cs:51.5,-1765) {C};
\end{axis}
\end{tikzpicture}%
}
\endgroup
\caption{Cross-method comparison. (a)~Running-best interaction energy over 25 iterations for Campaigns A--C, with horizontal baselines for DQN mean, PPO mean, and Optuna's best. (b)~Best interaction energy vs.\ unique PSMILES evaluated; error bars show $\pm 1\sigma$ over three RL seeds.}
\label{fig:results_panel}
\end{figure*}

\subsection{Cross-method comparison}

Figure~\ref{fig:comparison} plots best energy against unique PSMILES evaluated for all methods.
RL agents consume their budget $B$ efficiently (8~min) but achieve high regret because their restricted degrees of freedom (Table~\ref{tab:dof}) confine the search to 14 structures near the PEG seed.
Optuna reduces regret by 41\% relative to DQN's mean but at 15$\times$ the wall time and still without hypothesis-level reasoning.
Campaign~A achieves the lowest simple regret with 25+ unique structures; Campaigns~B and~C explore complementary chemical subspaces, collectively confirming that the agentic framework's additional degrees of freedom (the implicit acquisition function (Eq.~\ref{eq:implicit_acq}) and the discovery-world state update $u_\text{selective}$) translate into better regret under matched budgets.

\section{Analysis of Chemical Reasoning}
\label{sec:analysis}

Three autonomous campaigns on the same physics objective let us separate \emph{what} the LLM proposes from \emph{what} the force field and packing geometry allow.
Unlike RL and Optuna, the agent records hypotheses and open questions in the discovery world, producing an auditable trace of chemical intuition, dead ends, and revisions.

\subsection{Convergent design rules}

All three campaigns independently rank hydrogen-bond-dense motifs above ethers, esters, fluorinated, silicone, or purely aromatic backbones.
Campaign~A discovers a compact C2 scaffold with amine and multiple amides ($-2263$~kJ/mol); Campaign~B discovers $\alpha$-amino acid polymers whose side chains add imidazole (polyhistidine, $-1545$~kJ/mol) or primary amide (polyasparagine, $-1535$~kJ/mol); Campaign~C elevates polygalacturonic and polyglucuronic acids ($-1765$ and $-1574$~kJ/mol), linking high scores to polyol and carboxylate density.
This convergence across unrelated search policies supports a design rule: favor repeat units that present multiple NH and C=O (and, in Campaign~C, vicinal OH) contacts per monomer.

The discovery world captures this reasoning explicitly.
Campaign~B records \textit{``Hydrogen bonding capacity (not just polarity) correlates with interaction strength''} and \textit{``Polylysine has 2 amines per monomer but packmol failed, too sticky/bulky?''}
Campaign~C records \textit{``Polygalacturonic acid's multiple hydroxyl groups (3 per repeat) + aldehyde create strongest binding yet ($-$1765 kJ/mol)''} after iteration~7.
Campaign~A's logs refine the C2 backbone hypothesis by targeted falsification (C3 packing failure, thio-oxazolidinone collapse, ring-strain on four- and seven-membered lactams).
These traces show the agent using simulation outcomes to update qualitative models, producing high information gain $I_t$ (Eq.~\ref{eq:infogain}), not only to rank candidates.

\begin{figure*}[p]
\centering
\includegraphics[width=\textwidth]{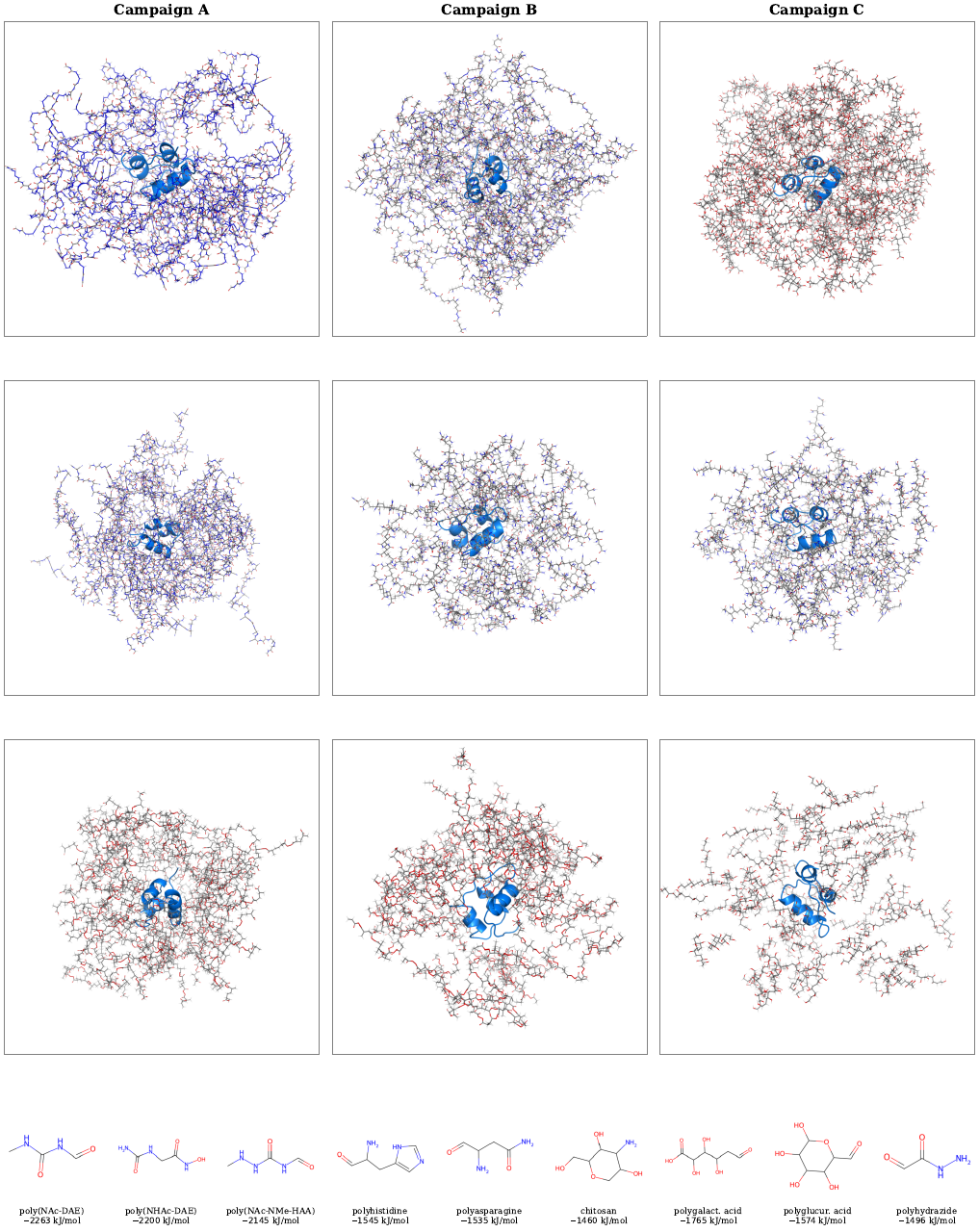}
\caption{Polymer--insulin complex structures and top-3 repeat-unit monomers from each agentic campaign. Top three rows: PyMOL renderings of representative polymer--insulin complexes (insulin shown as cartoon ribbon in blue; polymer as ball-and-stick), with each column corresponding to Campaign~A, B, or~C. Bottom row: 2D monomer depictions of the three highest-scoring candidates per campaign, labeled with abbreviated name and interaction energy.}
\label{fig:chemviz_panel}
\end{figure*}

\subsection{Structure--activity relationships}

\begin{table*}[t]
\centering
\caption{Top 10 candidates from Campaign~A (25 iterations), sorted by interaction energy. Three candidates exceed $-2100$~kJ/mol.}
\label{tab:agentic_top}
\small
\begin{tabular}{clcccl}
\toprule
\textbf{Rank} & \textbf{Polymer} & \textbf{PSMILES} & \textbf{$E_\text{int}$} & \textbf{Iter} & \textbf{Key motifs} \\
\midrule
1  & poly(N-acetyl-1,2-diaminoethane) & \texttt{\footnotesize [*]CNC(=O)NC([*])=O} & $-2263$ & 15 & amine + 2$\times$amide \\
2  & poly(N-hydroxyacetyl-diaminoethane) & \texttt{\footnotesize [*]C(NC(=O)N[*])C(=O)NO} & $-2200$ & 20 & hydroxamic acid \\
3  & poly(N-acetyl-hydrazinoacetamide) & \texttt{\footnotesize [*]CNNC(=O)NC([*])=O} & $-2145$ & 25 & hydrazide + amide \\
4  & poly(N-acetyl-biuret) & \texttt{\footnotesize [*]NCNC(=O)NC([*])=O} & $-1941$ & 22 & 3$\times$amide (biuret) \\
5  & poly(N-acetyl-1,2-diaminoethane) rep. & \texttt{\footnotesize [*]CNC(=O)NC([*])=O} & $-1909$ & 16 & replication \\
6  & poly(N-acetyl-oxalyldiamide) & \texttt{\footnotesize [*]CNC(=O)C(=O)NC([*])=O} & $-1866$ & 24 & 3$\times$amide (oxalyl) \\
7  & poly(N-vinyl lactamide) & \texttt{\footnotesize [*]CC([*])NC(=O)C(C)O} & $-1807$ & 12 & amide + hydroxyl \\
8  & poly(N,N'-diformylethylenediamine) & \texttt{\footnotesize [*]NC(=O)NC([*])=O} & $-1701$ & 14 & 2$\times$amide (urea) \\
9  & poly(N-acetyl-serinamide) & \texttt{\footnotesize [*]CC(O)CNC([*])=O} & $-1686$ & 23 & amide + hydroxyl \\
10 & poly(N-vinyl-2-oxazolidinone) & \texttt{\footnotesize [*]CC([*])N1CCOC1=O} & $-1665$ & 8  & cyclic urethane \\
\bottomrule
\end{tabular}
\end{table*}

\begin{table*}[t]
\centering
\caption{Top 10 candidates from Campaign~B, sorted by interaction energy.}
\label{tab:agentic_top_b}
\small
\begin{tabular}{clcccl}
\toprule
\textbf{Rank} & \textbf{Polymer} & \textbf{PSMILES} & \textbf{$E_\text{int}$} & \textbf{Iter} & \textbf{Key motifs} \\
\midrule
1  & polyhistidine & \texttt{\footnotesize [*]NC(Cc1cnc[nH]1)C([*])=O} & $-1545$ & 23 & amine, imidazole \\
2  & polyasparagine & \texttt{\footnotesize [*]NC(CC(N)=O)C([*])=O} & $-1535$ & 24 & amide (side chain) \\
3  & chitosan & \texttt{\footnotesize [*]OC1C([*])OC(CO)C(O)C1N} & $-1460$ & 1  & amine, hydroxyl, ether \\
4  & polyglutamine & \texttt{\footnotesize [*]NC(=O)CCC(N)C([*])=O} & $-1343$ & 25 & amide \\
5  & polyglutamic acid & \texttt{\footnotesize [*]CCC(NC([*])=O)C(=O)O} & $-1307$ & 3  & amine, acid \\
6  & polyaspartic acid & \texttt{\footnotesize [*]NC(CC(=O)O)C([*])=O} & $-1286$ & 2  & amine, acid \\
7  & PHEMA & \texttt{\footnotesize [*]CC([*])(C)C(=O)OCCO} & $-1129$ & 7  & ester, hydroxyl \\
8  & polyacrylamide & \texttt{\footnotesize [*]CC([*])C(N)=O} & $-1123$ & 2  & amide \\
9  & PLGA & \texttt{\footnotesize [*]OC(=O)COC(=O)C(C)[*]} & $-1047$ & 1  & ester (2) \\
10 & polyglycine & \texttt{\footnotesize [*]NCC([*])=O} & $-906$ & 17 & backbone amide \\
\bottomrule
\end{tabular}
\end{table*}

\begin{table*}[t]
\centering
\caption{Top 10 candidates from Campaign~C, sorted by interaction energy.}
\label{tab:agentic_top_c}
\small
\begin{tabular}{clcccl}
\toprule
\textbf{Rank} & \textbf{Polymer} & \textbf{PSMILES} & \textbf{$E_\text{int}$} & \textbf{Iter} & \textbf{Key motifs} \\
\midrule
1  & polygalacturonic acid & \texttt{\footnotesize [*]OC(=O)C(O)C(O)C(O)C([*])C=O} & $-1765$ & 7 & acid, polyol \\
2  & polyglucuronic acid & \texttt{\footnotesize [*]OC1C(O)C(O)OC(C([*])=O)C1O} & $-1574$ & 8 & acid, ether, polyol \\
3  & polyhydrazide & \texttt{\footnotesize [*]NNC(=O)C([*])=O} & $-1496$ & 9 & amide (2) \\
4  & polyglutaric acid & \texttt{\footnotesize [*]C(=O)CCC([*])=O} & $-1405$ & 10 & carbonyl \\
5  & polyglutamic acid & \texttt{\footnotesize [*]NC(CCC([*])=O)C(=O)O} & $-1385$ & 3 & amine, acid \\
6  & chitosan & \texttt{\footnotesize [*]OC1C([*])OC(CO)C(O)C1N} & $-1370$ & 1 & amine, hydroxyl \\
7  & polyacrylamide & \texttt{\footnotesize [*]CC([*])C(N)=O} & $-1256$ & 2 & amide \\
8  & polyaspartic acid & \texttt{\footnotesize [*]NC(CC(=O)O)C([*])=O} & $-1207$ & 4 & amine, acid \\
9  & polyitaconic acid & \texttt{\footnotesize [*]CC([*])(CC(=O)O)C(=O)O} & $-1144$ & 25 & acid (2) \\
10 & PHEMA & \texttt{\footnotesize [*]CC([*])(C)C(=O)OCCO} & $-1045$ & 5 & ester, hydroxyl \\
\bottomrule
\end{tabular}
\end{table*}

Poly(N-acetyl-1,2-diaminoethane) (\texttt{[*]CNC(=O)NC([*])=O}, $-2263$~kJ/mol, iteration~15; Table~\ref{tab:agentic_top}) defines the structural optimum.
Its C2 backbone carries a primary amine, a secondary amide, and a primary amide: three nitrogen atoms that provide multiple hydrogen-bond donors and complementary carbonyl acceptors.
This spacing matches the Asn, Gln, and Arg side chains exposed on the insulin B-chain helix (PDB 4F1C).
Replication in iteration~16 ($-1909$~kJ/mol) confirms robustness within the ${\sim}$15\% packing noise $\epsilon$ (Eq.~\ref{eq:oracle}); the two-evaluation mean of $-2086 \pm 250$~kJ/mol still exceeds Optuna's best by 10\%.

Campaign~A's systematic probing reveals several local structure--activity rules.
\emph{Backbone length:} the C3 variant failed to pack within 300~s, suggesting the C2 geometry is precisely complementary to insulin's surface.
\emph{Ring constraint:} ring-expanded analogs degrade progressively (5-membered oxazolidinone $-1665$, 6-membered oxazinone $-1239$, 7-membered packing failure), confirming that linear, flexible backbones outperform constrained cyclic scaffolds.
\emph{Heteroatom substitution:} thio-oxazolidinone scores $-830$~kJ/mol versus $-1665$~kJ/mol for the oxygen parent, a 50\% loss consistent with sulfur's weaker hydrogen-bond acceptance and larger van der Waals radius.
\emph{Hydroxyl cooperativity:} the hydroxamic acid variant ($-2200$~kJ/mol) demonstrates near-optimal binding when CONHOH is combined with a secondary amide; similarly, secondary hydroxyl placement outperforms primary ($-1807$ vs.\ $-1576$~kJ/mol).
\emph{Hydrazide motif:} introduced in the final iterations, secondary hydrazide ($-2145$~kJ/mol) enters the top~3, mirroring the primary/secondary amide trend.

\subsection{Divergent exploration strategies}

Each campaign adopts a distinct exploration strategy.
Campaign~A behaves like a medicinal chemistry cycle, systematically probing backbone length, ring size, heteroatom substitution, and N-substitution around the optimal scaffold.
Campaign~B resembles a formulation survey: batch evaluation of commodity polymers followed by a pivot (triggered when the discovery world records saturation) into biomimetic amino acid polymers.
Campaign~C stays in polyelectrolyte and polysaccharide chemistry, with a literature-driven emphasis on uronic acids.
Together they are complementary: one stress-tests local structure--activity relationships, another tests protein-like repeat units, and the third probes acidic polysaccharide motifs under the same screening oracle $f$.

\subsection{Hallucinations and physics as falsification}

LLM-driven name-to-structure pipelines can emit plausible-looking PSMILES that do not correspond to established repeat units.
In Campaign~B the agent labels a brominated diaryl sulfone as ``polysulfone''; real polysulfone is a bisphenol--ether--sulfone copolymer.
RDKit flags the name--graph inconsistency, and critically the Packmol-matrix evaluation returns $-404$~kJ/mol, ranking it among the weakest binders and preventing it from steering later iterations.

Steric and packing failures provide a second falsification layer.
Bulky aromatic side chains (tryptophan, tyrosine analogs), the C3 diamine backbone, and seven-membered cyclic urethanes all trigger Packmol timeouts: not ``bad energies'' but \emph{no valid bulk geometry}, which stops propagation of hypotheses that sound reasonable in natural language.
In the language of Eq.~\ref{eq:infogain}, these failures contribute high information gain $I_t$ because they sharply reduce the entropy of the agent's hypothesis set $\mathcal{H}_t$, pruning entire chemical families from future exploration.

Classical optimizers lack this safeguard: Optuna's peroxide-ester (\texttt{[*]OOC(=O)C([*])=O}, $-1171$~kJ/mol) contains a labile O--O bond that a domain expert would immediately reject, yet no mechanism within TPE prevents it from influencing subsequent trials.
None of the three agentic campaigns proposed peroxide linkers, illustrating how prompt-level chemistry and tool-mediated screening jointly constrain the search space.

\section{Discussion}

Taken together, the results connect a domain-scale need (cold-chain insulin) to a concrete polymer-design workflow and then to broader lessons about agentic optimization under physics oracles.
The agentic advantage documented in this work is structural, not statistical.
Campaign~A's 68\% regret reduction over RL and 19\% over Optuna does not stem from evaluating vastly more structures (115+ vs.\ 143 for Optuna); it stems from the additional degrees of freedom available to the LLM agent (Table~\ref{tab:dof}).
RL operates with $\mathcal{W}_t = \emptyset$ and a fixed mutation library $\{\alpha\}$, confining the search to ester and ether backbones near the PEG seed.
Optuna adds an explicit acquisition function but no hypothesis-level reasoning.
By conditioning on the full state $(\mathcal{D}_t, \mathcal{W}_t, \mathcal{H}_t)$, the LLM agent's implicit acquisition (Eq.~\ref{eq:implicit_acq}) accesses polymer families (heterocyclic urethanes, C2 diamide scaffolds, hydroxamic acids, hydrazides) that lie entirely outside the RL mutation library.
Campaigns~B and C confirm that even sparser trajectories (30--50+ evaluations) extract consistent design rules.
Information gain from falsified hypotheses (Eq.~\ref{eq:infogain}) prunes unpromising families early, reducing future exploration cost, a mechanism unavailable to methods without a world model.

These results should be interpreted in light of the screening oracle's limitations.
The Packmol-matrix protocol packs polymer chains around insulin in vacuum, minimizes, and computes a single-point interaction energy.
It does not sample conformational ensembles, account for explicit solvent, or compute binding free energies.
Replication reveals ${\sim}$15\% stochastic variability (iteration~15: $-2263$~kJ/mol; iteration~16: $-1909$~kJ/mol), so energy differences below ${\sim}$200~kJ/mol should be interpreted cautiously; the ${\sim}$900~kJ/mol gap between the agentic best and RL remains robust.
This oracle noise $\epsilon$ (Eq.~\ref{eq:oracle}) also limits the precision of the implicit acquisition.
Improving $f$ through NPT dynamics with the UMA universal force field~\cite{wood2025uma} or free-energy perturbation would sharpen the objective surface and likely amplify the agentic advantage.

The energy metric itself does not capture synthetic accessibility or thermal stability.
Optuna's peroxide-ester ($-1171$~kJ/mol) illustrates the risk of energy-only optimization: the labile O--O bond would be immediately flagged by any domain expert.
The agentic workflow avoids such proposals through implicit chemical knowledge, but systematic integration of synthesizability scores into the objective remains planned work.

A practical advantage of the platform is its compatibility with local and edge deployment.
The entire science stage (OpenMM, Packmol, RDKit, and the RL baselines) ran on CPU hosts in this study; the LLM issues tool calls and updates JSON state.
This split aligns with emerging on-device and mobile-edge LLM systems that prioritize latency, cost, and data localization~\cite{xu2024ondevice,qu2024mobileedge}, and with local serving stacks such as Ollama~\cite{ollama2023} that make open-weight models practical on workstations.

Kosmos~\cite{kosmos2025} executes ${\sim}$42,000 lines of code and reads 1,500 papers per 12-hour run, operating on a broad multi-domain $\mathcal{X}$ but without a physics oracle $f$.
Our platform operates at a smaller scale yet provides two capabilities Kosmos lacks: physics-grounded screening through explicit simulation and iterative structure--activity reasoning that produces falsifiable hypotheses.
The probabilistic-chain formalism~\cite{stephens2025mathframing} provides a common language for comparing such architectures: differences in $\mathcal{F}$, $u$, and $\{\alpha\}$ map directly to differences in achievable regret.

\section{Conclusions}

We have presented an agentic discovery platform that couples LLM-driven reasoning with physics-based molecular simulation to search the polymer space for insulin-stabilizing transdermal matrices.
Three independent campaigns converge on a single design rule: dense hydrogen-bond donors and acceptors per repeat unit, while physics validation rejects hallucinated structures before they propagate.
Falsified hypotheses produce high information gain (Eq.~\ref{eq:infogain}), accelerating convergence by pruning entire chemical families.

The formal framework developed here (probability chain, implicit acquisition, simple regret) reveals that the agentic advantage is not about more evaluations but about more degrees of freedom in the optimization: the LLM agent conditions on a structured world model and a full MCP tool set, whereas RL and Bayesian optimization operate on strict subsets of this space (Table~\ref{tab:dof}).
This analysis, grounded in the formalism of Stephens and Salawu~\cite{stephens2025mathframing} and budget-constrained regret theory~\cite{shahriari2016bayesopt}, provides a principled basis for comparing future agentic strategies.

Because the simulation stage is CPU-bound, the entire workflow runs on commodity hardware, compatible with local and edge LLM deployments~\cite{xu2024ondevice,qu2024mobileedge} and local serving stacks~\cite{ollama2023}, making it accessible to resource-constrained laboratories.
Replacing the current single-point screening oracle with higher-fidelity molecular dynamics (explicit solvent, free-energy perturbation, universal force fields~\cite{wood2025uma}) would tighten estimates of interaction strength and is the natural next step.
Beyond insulin, the same platform architecture generalizes to any protein-stabilization or biologics-formulation problem (vaccines, enzyme therapeutics, antibody storage) wherever a computationally tractable screening oracle exists.
Top candidates from the campaigns reported here merit experimental follow-up with stability and release assays.

Code, benchmark scripts, and session folders are available in the project repository.

\section{Acknowledgments}

We acknowledge the developers of OpenMM, RDKit, Packmol, Stable-Baselines3, Optuna, and the Open Force Field Initiative for providing the foundational software infrastructure.
We thank the Semantic Scholar team at the Allen Institute for AI for access to their academic API.
The author also acknowledges MRH Scientific for computational resources.

\bibliographystyle{plainnat}
\bibliography{references}

\appendix

\section{Benchmark Algorithms}
\label{app:algorithms}

Algorithms~\ref{alg:rl} and~\ref{alg:optuna} formalize the two non-agentic baselines using the notation from \S\ref{sec:formulation}.
Both operate on a strict subset of the degrees of freedom available to the agentic workflow (Table~\ref{tab:dof}): neither maintains a hypothesis state $\mathcal{W}_t$ or uses a structured state update $u_\text{selective}$.

\begin{algorithm}[H]
\caption{RL Polymer Discovery (DQN / PPO)}\label{alg:rl}
\small
\begin{algorithmic}[1]
\Require seed PSMILES $x_0$, oracle $f$, mutation library $\{\alpha\}$, episodes $E$, steps $K$
\State $\mathcal{F} \gets$ initialize policy network \Comment{fixed $\mathcal{F}$}
\State $\mathcal{W}_t \gets \emptyset$ for all $t$ \Comment{no world model}
\For{episode $e = 1, \ldots, E$}
  \State $x \gets x_0$
  \For{step $k = 1, \ldots, K$}
    \State $\alpha_k \gets \mathcal{F}(x)$ \Comment{select mutation from library}
    \State $x' \gets \alpha_k(x)$ \Comment{apply mutation}
    \State $y \gets f(x')$ \Comment{Eq.~\ref{eq:oracle}: oracle query}
    \State reward $\gets -y$
    \State update $\mathcal{F}$ with $(x, \alpha_k, \text{reward}, x')$
    \State $x \gets x'$
  \EndFor
\EndFor
\State \Return $\arg\min_{x \in \mathcal{D}} f(x)$
\end{algorithmic}
\end{algorithm}

\begin{algorithm}[H]
\caption{Optuna TPE Polymer Search}\label{alg:optuna}
\small
\begin{algorithmic}[1]
\Require seed PSMILES $x_0$, oracle $f$, mutation library $\{\alpha\}$, trials $T$, candidates/trial $C$
\State $\mathcal{W}_t \gets \emptyset$ for all $t$ \Comment{no hypothesis state}
\For{trial $\tau = 1, \ldots, T$}
  \State $(\text{seed}, \text{frac}) \gets \text{TPE sample}(\mathcal{D}_{\tau-1})$ \Comment{explicit acquisition}
  \For{$j = 1, \ldots, C$}
    \State $\alpha_j \gets$ select mutation from $\{\alpha\}$ using seed
    \State $x_j \gets \alpha_j(x_0,\, \text{frac})$
    \State $y_j \gets f(x_j)$ \Comment{Eq.~\ref{eq:oracle}: oracle query}
    \State $\mathcal{D}_\tau \gets \mathcal{D}_\tau \cup \{(x_j, y_j)\}$
  \EndFor
  \State score $\gets$ composite$(y_1,\ldots,y_C,\, |\{x_j\}|)$
  \State report score to TPE sampler
\EndFor
\State \Return $\arg\min_{x \in \mathcal{D}_T} f(x)$
\end{algorithmic}
\end{algorithm}

\section{Agent Reasoning Excerpts}
\label{app:reasoning}

The following excerpts illustrate how the agentic discovery loop operates in practice, annotated with the formal notation from \S\ref{sec:formulation}.
Each excerpt corresponds to a specific phase of the ReAct cycle (Eq.~\ref{eq:react}): the agent generates a thought $t_i$, selects an action $\alpha_i(x_i)$, receives an observation $o_i$, and updates the world state $\mathcal{W}_t$.

\paragraph{Excerpt 1: Initial screening and hypothesis formation (Campaign~B, iteration~1).}
The agent screens six commodity polymers and forms its first qualitative hypotheses from the energy ranking.

\begin{agentthought}
\scriptsize
\textbf{Query:} ``hydrogel polymer insulin transdermal patch stabilization room temperature.''
I will evaluate a panel of well-known biocompatible polymers against insulin.
PEG is the most common hydrogel backbone; chitosan is widely used in drug delivery for its amine and hydroxyl groups; PLGA and PLA are FDA-approved biodegradable polyesters; PCL is a slow-degrading polyester; PVA provides dense hydroxyl groups.
I need to rank them by interaction energy and identify which functional groups drive stabilization.
\end{agentthought}

\begin{toolcall}[\texttt{evaluate\_candidates}($x_1,\ldots,x_6$): oracle queries]
\scriptsize
PEG: $-549$~kJ/mol. PLGA: $-1047$~kJ/mol. Chitosan: $-1460$~kJ/mol (\textbf{best}).
PCL: $-712$~kJ/mol. PVA: $-698$~kJ/mol. PLA: $-770$~kJ/mol.
\end{toolcall}

\begin{agentthought}
\scriptsize
Chitosan dominates with $-1460$~kJ/mol, driven by its dense amine + hydroxyl groups.
PLGA ($-1047$) has dual esters (polar but no NH donors). PEG ($-549$) has only ether oxygens: weak acceptors, zero donors.
Ranking: amine/hydroxyl $\gg$ ester $>$ ether.

\textbf{Hypothesis H1:} Chitosan's amine + hydroxyl groups provide strongest hydrogen bonding to insulin.
\textbf{Hypothesis H3:} Hydrogen bonding capacity (not just polarity) correlates with interaction strength.
\textbf{Open question Q2:} Can we improve on chitosan by adding more hydroxyl/amine groups or combining with other polymers?
\end{agentthought}

\begin{toolcall}[\texttt{patch\_discovery\_world}($x_i$): update $\mathcal{W}_t$]
\scriptsize
\texttt{\{"hypotheses": [\{"id":"H1", ...\}, \{"id":"H3", "text":"Hydrogen bonding capacity (not just polarity) correlates with interaction strength"\}], "open\_questions": ["Q2: Can we improve on chitosan by adding more hydroxyl/amine groups?"]\}}
\end{toolcall}

\noindent This thought $t_i$ (Eq.~\ref{eq:react}) conditions all subsequent tool calls: the agent prioritizes amide- and amine-rich candidates over ester-only backbones.
Under RL or Optuna, this qualitative insight would have no mechanism to influence future queries ($\mathcal{W}_t = \emptyset$).

\paragraph{Excerpt 2: Hallucination and physics-based correction (Campaign~B, iteration~6).}
The agent proposes ``polysulfone'' from the literature, but PubChem returns a brominated diaryl sulfone PSMILES instead of the expected bisphenol A polysulfone, a name--structure mismatch that the oracle catches.

\begin{agentthought}
\scriptsize
\textbf{Query:} ``polysulfone from literature.''
Literature suggests polysulfone for thermal stability.
Polysulfone should be a rigid aromatic backbone with sulfone groups (thermally stable, used in membranes).
Converting ``polysulfone'' to PSMILES via PubChem fallback\ldots
\end{agentthought}

\begin{toolcall}[\texttt{evaluate\_candidates}($x_i$): oracle query $f(x_i)$]
\scriptsize
\texttt{[*]C(Br)(c1ccccc1)S(=O)(=O)c1ccc(C)cc1[*]} $\to$ $E_\text{int} = -404$~kJ/mol.
\texttt{name\_consistency: false}. RDKit flags: bromine substituent absent from ``polysulfone'' definition; Tanimoto to expected structure: 0.31.
\end{toolcall}

\begin{observation}
\scriptsize
Result: $f(x_i) = -404$~kJ/mol, ranked among the weakest binders, worse than PEG.
The PubChem fallback resolved ``polysulfone'' to a small-molecule brominated diaryl sulfone rather than the bisphenol~A-based polymer.
\textbf{High $I_t$} (Eq.~\ref{eq:infogain}): aromatic sulfone motifs are pruned from $\mathcal{H}_t$.
Updated $\mathcal{W}_t$: ``Aromatic-rich polymers (polysulfone) show very poor interaction $-404$~kJ/mol.''
\textbf{Hypothesis H12:} Aromatic polymers consistently underperform due to lack of polar groups for H-bonding.
\end{observation}

\noindent The physics oracle rejects the hallucinated structure \emph{before} it can influence the next hypothesis.
Under RL or Optuna, a $-404$~kJ/mol result would simply be a low-reward data point; here the agent extracts a general lesson (aromatic backbones are weak binders) that prunes an entire chemical family from future exploration.

\paragraph{Excerpt 3: Iterative refinement and breakthrough (Campaign~A, iteration~12).}
Building on 11 prior iterations, the agent reasons from accumulated SAR to design candidates that combine the best-performing functional groups.

\begin{agentthought}
\scriptsize
\textbf{Query:} ``N-vinyl lactamide primary amide hydroxyl polymer insulin stabilization biomedical.''
Current best is poly(N-vinyl-2-oxazolidinone) at $-1665$~kJ/mol (iteration~8, 5-membered ring with amide + ether).
I should test: (1)~primary amide NH$_2$ + secondary hydroxyl (the lactamide motif); (2)~dual amide with ethylene spacer; (3)~thio-oxazolidinone (replace O with S in the ring); (4)~extended serinamide with dual hydroxyls.
Compact structures with amide directly attached to the alpha carbon gave best packing so far.
\end{agentthought}

\begin{toolcall}[\texttt{evaluate\_candidates}($x_1,\ldots,x_7$): oracle queries]
\scriptsize
\textbf{poly(N-vinyl lactamide):} \texttt{[*]CC([*])NC(=O)C(C)O} $\to$ $E_\text{int} = -1807$~kJ/mol. \textbf{NEW BEST} ($+143$ over oxazolidinone).
poly(N-acetyl-N-vinylglycinamide): $-1647$~kJ/mol (new \#2).
poly(N-vinyl-2-oxazolidinethione) (S$\to$O swap): $-830$~kJ/mol (\textbf{catastrophic}).
poly(2-oxopropyl acrylate): $-1529$~kJ/mol.
poly(N-acetylserinamide extended): \texttt{failed\_timeout}.
poly(N-acetyl-N-methyl acrylamide): \texttt{failed\_timeout}.
\end{toolcall}

\begin{observation}
\scriptsize
\textbf{Key SAR:} primary amide (NH$_2$) + secondary hydroxyl is the optimal combination ($-1807$~kJ/mol).
Thio-oxazolidinone ($-830$) confirms O $\gg$ S for H-bonding: larger atomic radius and weaker C=S acceptor.
Dual hydroxyls and dual tertiary amides both failed to pack; bulkier side chains cause Packmol timeouts.
Agent updates $\mathcal{W}_{12}$: ``primary amide NH$_2$ $\gg$ secondary amide $\gg$ tertiary amide for H-bond donation.''
\end{observation}

\paragraph{Excerpt 4: Structure--property reasoning leads to all-time best (Campaign~A, iteration~15).}
The agent synthesizes 14 iterations of accumulated SAR to design the optimal scaffold.

\begin{agentthought}
\scriptsize
\textbf{Query:} ``N-acetyl-1,2-diaminoethane primary amine secondary amide polymer insulin stabilization.''
\textbf{Reasoning from $\mathcal{W}_{14}$:}
Iteration~12 showed primary amide NH$_2$ + hydroxyl = $-1807$.
Iteration~14 showed dual primary amides (urea-type, diformylethylenediamine) = $-1701$.
Critical finding: primary amide NH donor $\gg$ secondary amide (NCH$_3$).
Diacetyl variant (secondary amide): $-1327$ vs.\ diformyl (primary amide): $-1701$ = 374~kJ/mol gap.
\textbf{Design hypothesis:} A C2 backbone (ethylenediamine) with primary amine + secondary amide should maximize H-bond donor density while keeping the repeat unit compact.
Testing 8 candidates: poly(N-acetyl-1,2-diaminoethane), N-formyl variants, oxalyldiamide, lactyl glycinamide, and C3 spacer controls.
\end{agentthought}

\begin{toolcall}[\texttt{evaluate\_candidates}($x_1,\ldots,x_8$): oracle queries]
\scriptsize
\textbf{poly(N-acetyl-1,2-diaminoethane):} \texttt{[*]CNC(=O)NC([*])=O} $\to$ $E_\text{int} = -2263$~kJ/mol.
\textbf{NEW ALL-TIME BEST} by 456~kJ/mol over lactamide.
poly(N-formyl-1,3-diaminopropane): $-1744$ (new \#3).
poly(N,N'-oxalyldiamide) (tertiary amides only): $-1637$ (moderate).
poly(N,N'-diacetyl-1,3-diaminopropane): $-1584$ (confirms C2 $>$ C3).
poly(N-formyl-N-hydroxy-ethylenediamine): $-1451$.
poly(N-lactyl glycinamide): $-1412$.
poly(N,N'-diformylhydrazine): $-1116$.
\end{toolcall}

\begin{observation}
\scriptsize
$f(x_i) = -2263$~kJ/mol. Simple regret $r_B$ (Eq.~\ref{eq:regret}) reduced by 33\% in a single iteration.
The implicit acquisition $\pi_\text{LLM}(s_t)$ (Eq.~\ref{eq:implicit_acq}), conditioned on accumulated SAR from $\mathcal{W}_{14}$, selected a candidate outside both the RL mutation library and Optuna's search space.
\textbf{Key SAR update:} amine (NH) + secondary amide in C2 backbone = optimal; chain length critical (C2 $>$ C3 $>$ C4);
tertiary amides weaker than primary/secondary.
Agent records: ``MASSIVE BREAKTHROUGH: amine + secondary amide + primary amide motif in C2 backbone is the optimal geometry.''
\end{observation}

\paragraph{Excerpt 5: Replication and noise estimation (Campaign~A, iteration~16).}
Rather than immediately pursuing new scaffolds, the agent replicates the best candidate to estimate oracle noise $\epsilon$ (Eq.~\ref{eq:oracle}).

\begin{agentthought}
\scriptsize
\textbf{Query:} ``N-acetyl-1,2-diaminoethane chain length variation insulin stabilization biomedical.''
The $-2263$~kJ/mol result may be an outlier.
I need to replicate poly(N-acetyl-1,2-diaminoethane) multiple times to estimate the true performance and assess stochastic packing variability.
I will also test chain-length controls: C4 spacer (1,4-diaminobutane) and shorter variants.
\end{agentthought}

\begin{toolcall}[\texttt{evaluate\_candidates}($x_1,\ldots,x_6$): oracle queries]
\scriptsize
poly(N-acetyl-1,2-diaminoethane) replicate~1: $-1817$~kJ/mol.
poly(N-acetyl-1,2-diaminoethane) replicate~2: $-1909$~kJ/mol.
poly(N-acetyl-1,2-diaminoethane) replicate~3: $-1905$~kJ/mol.
poly(N,N'-diacetyl-1,4-diaminobutane) (C4): \texttt{failed\_timeout}.
poly(3-aminopropyl urea): $-1051$~kJ/mol.
poly(2-aminoethyl formamide): $-1189$~kJ/mol.
\end{toolcall}

\begin{observation}
\scriptsize
Three replicates: $-1817$, $-1909$, $-1905$~kJ/mol (mean $\approx -1877$, $\sigma \approx 51$).
Original $-2263$ is ${\sim}7.5\sigma$ above the mean, likely a favorable packing outlier.
The true performance is $\approx -1877$~kJ/mol, still the all-time best.
C4 spacer failed to pack: confirms C2 is the optimal backbone length.
Agent updates $\mathcal{W}_{16}$: ``${\sim}$15\% stochastic variability from Packmol packing; energy differences $< 200$~kJ/mol are within noise.''
\end{observation}

\noindent This self-initiated replication (a form of epistemic caution encoded in $t_i$) has no analog in RL or Optuna.
The agent uses repeated oracle queries not to find new candidates but to sharpen its estimate of $f(x^\star)$ and calibrate future comparisons.

\paragraph{Excerpt 6: Saturation detection and biomimetic pivot (Campaign~B, iterations~8--10).}
After 10 iterations probing commodity polymers, the agent detects stagnation and formulates a new search strategy.

\begin{agentthought}
\scriptsize
\textbf{Iteration~8 query:} ``mutation attempt.''
Iterations 4--8 show diminishing returns.
Current leaderboard stalled at chitosan ($-1460$).
New candidates: poly(propylene imine) at $-598$; simple amine-only polymers underperform, need multiple H-bond groups.
PVDF at $-331$; fluorinated polymers very poor.
PDMAEMA failed to pack.
\textbf{Open question Q9:} ``No new high performers found for 4+ iterations; candidate pool saturated.''
\textbf{Open question Q10:} ``Need new approaches: polymer blends, co-polymers, or different search queries.''
\end{agentthought}

\begin{agentthought}
\scriptsize
\textbf{Iteration~17 query:} ``zwitterionic and wild chemistries.''
\textbf{Pivot reasoning:} If proteins stabilize proteins (chaperones, albumin excipients), then poly(amino acid) homopolymers (polyglycine, polyleucine, polyhistidine) may provide biomimetic stabilization.
They have dense amide backbones with functional side chains.
Polyglycine: simplest peptide backbone.
\end{agentthought}

\begin{toolcall}[\texttt{evaluate\_candidates}: amino acid polymers]
\scriptsize
Polyglycine: $-906$~kJ/mol. Polyleucine: $-879$~kJ/mol.
Both improve over PVP ($-845$) but trail chitosan.
PVDF: $-331$~kJ/mol. PDMAEMA: \texttt{failed\_timeout}.
\end{toolcall}

\begin{toolcall}[\texttt{patch\_discovery\_world}($x_i$): update $\mathcal{W}_t$]
\scriptsize
\texttt{\{"hypotheses": [\{"id":"H8", "text":"Proteins stabilize proteins -- amino acid polymers may provide biomimetic stabilization", "status":"open"\}], "open\_questions": ["Can amino acid homopolymers pack around insulin?", "Which side chains contribute most?"]\}}
\end{toolcall}

\noindent This pivot, impossible under RL's fixed mutation library ($\{\alpha\}$) or Optuna's parametric search, exploits the LLM's broader action space and hypothesis-driven reasoning.
It leads to polyhistidine ($-1545$~kJ/mol, Excerpt~7) and polyasparagine ($-1535$~kJ/mol), validating the biomimetic hypothesis.

\paragraph{Excerpt 7: Amino acid breakthrough (Campaign~B, iteration~23).}
Six iterations after the pivot, the agent reaches the imidazole-containing amino acid polymer.

\begin{agentthought}
\scriptsize
\textbf{Query:} ``histidine polymer.''
Polyglycine ($-906$) and polyleucine ($-879$) are moderate.
Histidine's imidazole ring has two nitrogen atoms: one is a strong H-bond donor (NH), the other a strong acceptor (lone pair on N).
This dual role should increase H-bonding density per repeat unit beyond what glycine or leucine can provide.
If the biomimetic hypothesis (H8) is correct, histidine should outperform the simple amino acid backbones.
\end{agentthought}

\begin{toolcall}[\texttt{evaluate\_candidates}($x_i$): oracle query $f(x_i)$]
\scriptsize
\texttt{[*]NC(Cc1cnc[nH]1)C([*])=O} $\to$ $E_\text{int} = -1545$~kJ/mol.
\textbf{NEW BEST} for Campaign~B; beats chitosan by 85~kJ/mol.
\end{toolcall}

\begin{observation}
\scriptsize
\textbf{Hypothesis H8 validated:} amino acid polymers can match or exceed polysaccharides.
The imidazole side chain provides the additional H-bonding capacity that glycine and leucine lacked.
Agent records: ``Major breakthrough! Polyhistidine with imidazole ring beats chitosan by 85~kJ/mol.''
$\mathcal{W}_{23}$ now prioritizes imidazole-containing and amide-rich side chains for remaining iterations.
\end{observation}

\paragraph{Excerpt 8: Information gain from systematic failure (Campaign~A, iteration~21).}
The agent tests ring-expanded analogs of the early-campaign oxazolidinone hit, all of which underperform, yielding high information gain from falsification.

\begin{agentthought}
\scriptsize
\textbf{Query:} ``oxazolidinone methyl derivatives thiazolidinone hydroxy-oxazolidinone ring-expanded lactams.''
The 5-membered oxazolidinone scored $-1665$ at iteration~8.
Can ring modifications improve it?
Testing: gem-dimethyl-oxazolidinone, thiazolidinone (S replaces O), 4-hydroxy-2-oxazinone (6-membered), 2-piperidinone (6-membered saturated), and hydroxyethyl barbiturate.
\textbf{Hypothesis:} ring expansion may reduce strain and improve packing, or larger rings may provide more H-bonding sites.
\end{agentthought}

\begin{toolcall}[\texttt{evaluate\_candidates}($x_1,\ldots,x_6$): oracle queries]
\scriptsize
gem-dimethyl-oxazolidinone: $-1476$~kJ/mol. Thiazolidinone: $-1352$. Hydroxyethyl barbiturate: $-1421$.
4-hydroxy-2-oxazinone (6-membered): $-1102$. 2-Piperidinone (6-membered): $-1042$.
4-hydroxy-2-oxazolidinone: $-1307$.
\textbf{All} $< -1476$; none approaches the linear C2 amide scaffold ($-2263$).
\end{toolcall}

\begin{observation}
\scriptsize
Ring expansion progressively hurts: 5-membered ($-1665$) $>$ 6-membered ($-1102$) $>$ piperidinone ($-1042$).
\textbf{High $I_t$} (Eq.~\ref{eq:infogain}): the entire ``ring-based scaffold'' family is falsified.
This prunes a large subset of $\mathcal{X}$ from future exploration, reducing $H(\mathcal{H}_t)$ and focusing subsequent queries on linear amide variants.
Agent records: ``Ring expansion hurts (5$>$6$>$7-membered). Ring-based scaffolds inferior to linear C2 amide scaffolds.''
Under RL or Optuna, these six evaluations would be six low-reward data points with no structural consequence for future search.
\end{observation}

\paragraph{Excerpt 9: Final convergence and hydrazide discovery (Campaign~A, iteration~25).}
In the final iteration, the agent explores hydrazide isosteres of the C2 scaffold, confirming convergence and discovering a new top-5 candidate.

\begin{agentthought}
\scriptsize
\textbf{Query:} ``hydrazide N-methyl-hydrazide amino-acetonitrile cyanoacetamide final frontier insulin stabilization.''
This is the final iteration. The C2 scaffold with amine + secondary amide dominates.
\textbf{Can hydrazide (NHNH) replace amide (NHC=O) as H-bond donor?}
Secondary hydrazide adds an extra NH without disrupting the C2 geometry.
Also testing: nitrile group (strong dipole, weak H-bond donor), primary hydrazide, oxalyldiamide variant.
Goal: confirm convergence and probe the frontier of the amide isostere space.
\end{agentthought}

\begin{toolcall}[\texttt{evaluate\_candidates}($x_1,\ldots,x_5$): oracle queries]
\scriptsize
\textbf{poly(N-acetyl-N-methyl-hydrazinoacetamide):} \texttt{[*]CNNC(=O)NC([*])=O} $\to$ $E_\text{int} = -2145$~kJ/mol. \textbf{Enters all-time top~3.}
poly(N-acetyl-hydrazinoacetamide): $-1698$~kJ/mol (good).
poly(N-acetyl-N-hydrazinopropanamide): $-1667$~kJ/mol.
poly(N-acetyl-2-cyanoacetamide): $-1460$~kJ/mol (nitrile is weak).
poly(N-acetyl-oxalyldiamide): $-1513$~kJ/mol.
\end{toolcall}

\begin{observation}
\scriptsize
Secondary hydrazide is a strong H-bond donor: $-2145$~kJ/mol confirms it as a viable amide isostere.
Nitrile ($-1460$) is weak: dipole alone is insufficient without NH donors.
\textbf{Convergence confirmed:} the N-acetyl-1,2-diaminoethane scaffold and its hydrazide variant dominate.
All top-5 candidates share the motif: C2 backbone + amine + secondary amide/hydrazide + primary amide.

Agent's final SAR summary (10 rules in $\mathcal{W}_{25}$): (1)~C2 backbone optimal; (2)~primary amide NH $\gg$ secondary $\gg$ tertiary; (3)~secondary alcohol $>$ primary $\gg$ tertiary; (4)~hydroxyl + amide is additive; (5)~ring expansion hurts (5$>$6$>$7); (6)~O $\gg$ S in all comparisons; (7)~methyl on amide N is optimal; (8)~hydrazide is a viable isostere; (9)~bulkier side chains cause packing failures; (10)~multiple amides (3) provide good stabilization but cannot match C2 geometry.
\end{observation}

\end{document}